\documentclass[aps,nopacs,nokeys]{revtex4}
\usepackage{amsmath, amstext, amssymb, amsthm}
\newcommand{\bra}[1]{\langle#1|}
\newcommand{\ket}[1]{|#1\rangle}

\newcommand{\ketbra}[2]{|#1\rangle\!\langle#2|}
\newcommand{\pt}{^{\Gamma}}
\newcommand{\eref}[1]{(\ref{#1})}

\newcommand{\Perr}[1]{P_{\rm err}^{\rm #1}}
\newcommand{\bias}[1]{B^{\rm #1}}
\newcommand{\CI}[1]{\xi^{\rm #1}}
\def\tr{\mathinner{\mathrm{Tr}}}

\def\ALL{\mathinner{\mathrm{ALL}}}
\def\SEP{\mathinner{\mathrm{SEP}}}
\def\PPT{\mathinner{\mathrm{PPT}}}
\def\LOCC{\mathinner{\mathrm{LOCC}}}
\def\ox{\otimes}
\def\OX{\bigotimes}
\def\a{\alpha}
\def\s{\sigma}
\def\1{\openone}

\newtheorem{theorem}{Theorem}

\newtheorem{remark}[theorem]{Remark}
\newtheorem{proposition}[theorem]{Proposition}
\newtheorem{corollary}[theorem]{Corollary}
\newtheorem{conjecture}[theorem]{Conjecture}

\begin{document}
	\title{On the Chernoff distance for asymptotic\protect\\ LOCC discrimination of bipartite quantum states}
	\author{William Matthews}
	\email{william.matthews@bris.ac.uk}
	\affiliation{Department of Mathematics, University of Bristol, Bristol BS8 1TW, U.K.}
	\author{Andreas Winter}
	\email{a.j.winter@bris.ac.uk}
	\affiliation{Department of Mathematics, University of Bristol, Bristol BS8 1TW, U.K.}
	\affiliation{Centre for Quantum Technologies, National University of Singapore, 2 Science Drive 3, Singapore 117542}
	\date{22 October 2007}
	
\begin{abstract}
	Motivated by the recent discovery of a quantum Chernoff theorem for asymptotic state discrimination, we investigate the distinguishability of two bipartite mixed states under the constraint of local operations and classical communication (LOCC), in the limit of many copies. While for two pure states a result of Walgate \emph{et al.} shows that LOCC is just as powerful as global measurements, data hiding states (DiVincenzo \emph{et al.}) show that locality can impose severe restrictions on the distinguishability of even orthogonal states.
	Here we determine the optimal error probability and measurement to discriminate many copies of particular data hiding states (extremal $d\times d$ Werner states) by a linear programming approach. Surprisingly, the single-copy optimal measurement remains optimal for $n$ copies, in the sense that the best strategy is measuring each copy separately, followed by a simple classical decision rule. We also put a lower bound on the bias with which states can be distinguished by separable operations.
\end{abstract}
	
\maketitle

\section{Introduction}
\label{sec:intro}


The non-classical nature of information represented in states of a bipartite quantum system is strikingly evident in the fact that, even allowing the experimenters (Alice and Bob) holding each of the subsystems to use local operations and \emph{classical} communication (LOCC) freely, they cannot access the information as well as if they were in the same lab or could exchange quantum states. Thus, there is a specifically quantum obstruction to the distributed analysis of data and investigating this obstruction is a way of obtaining an understanding of the quantum nature of information.

The problem of LOCC discrimination of two or more states, has recently attracted quite considerable attention~\cite{walgate,WC1,WC2,WC3,WC4,WC5,WC6,WC7,WC8,WC9,virmani,king,nathanson} and what can be said at the very least that it is difficult.
In the simplest example, the experimenters are given one of two states at random according to some probability distribution and their task is to unambiguously determine which state they have with the smallest possible error probability. Throughout this paper we'll use $\Perr{X}\left(\rho_1,\rho_2; p\right)$ to denote the minimum error with which the states $\rho_1$ and $\rho_2$, with prior probabilities $p$ and $1-p$ respectively, can be distinguished by a POVM that can be implemented by operations in the class $X$. It will sometimes be convenient to refer to the optimal bias (over random guessing) instead of the optimal probability. This we define, as usual, by
\begin{equation}
	\bias{X}= 1 - 2 \Perr{X}.
\end{equation}
In this work we will talk about the well known classes of PPT-preserving (PPT) operations, separable (SEP) operations \cite{rains} and local operations with classical communication (LOCC), which obey the strict inclusions~\cite{nonlocality-w-o-ent}
\begin{equation}
	\LOCC \subset \SEP \subset \PPT \subset \ALL,
\end{equation}
where $\ALL$ simply denotes the set of all possible global operations. Briefly, the POVMs which can be implemented by operations in these different classes can be characterized as follows: An LOCC POVM is one which can be implemented as a multi-round process where each round consists of a partial measurement of one party, which can depend on previously generated classical messages, and whose result is broadcast; A POVM is in SEP if and only if its elements can be written as positive linear combinations of product operators; A POVM can be implemented by PPT operations if and only if its constituent operators have positive partial transpose. The inclusion structure immediately implies the ordering
\begin{equation}
\label{errorinequalities}
	\Perr{LOCC} \geq \Perr{SEP} \geq \Perr{PPT} \geq \Perr{ALL} = \frac{1}{2} - \frac{1}{2} \left\| p \rho_1 - (1-p) \rho_2 \right\|_1.
\end{equation}
The final equality is the classic result of Helstrom and Holevo \cite{helstrom}. A similar closed form expression does not seem to exist for $\Perr{LOCC}$ or any of the other bipartite
$\Perr{X}$.

Motivated by the recent development of a quantum Chernoff theorem \cite{q-chernoff}, we are interested here in the asymptotic behaviour of the quantity $\Perr{X}\left(\rho_1^{\ox n},\rho_2^{\ox n};p\right)$ as the number of copies, $n$, goes to infinity. We can define the \emph{Chernoff distance} with respect to a class of operations $X$, between the states $\rho_1$ and $\rho_2$ by
\begin{equation}
\label{eq:CI-defn}
	\CI{X}\left(\rho_1,\rho_2\right) = \lim_{n\to\infty} -\frac{1}{n} \log \Perr{X}\left(\rho_1^{\ox n},\rho_2^{\ox n};p\right).
\end{equation}
(We note that the Chernoff distance is not strictly a distance since it does not obey the triangle inequality and that it is independent of the prior probabilities as long as they are both non-zero.)

In \cite{q-chernoff}, it was determined that the (unconstrained) quantum Chernoff distance $\CI{ALL}\left(\rho_1,\rho_2\right)$ is given by the formula (note the independence of $p$):
\begin{equation}
	\CI{ALL}\left(\rho_1,\rho_2\right) = -\min_{0\leq s \leq 1} \log \tr \rho_1^{1-s} \rho_2^{s}.
\end{equation}
This is a pleasantly straightforward generalisation of the classical Chernoff theorem for probability distributions, where for probability distribution vectors $p$ and $q$
\begin{equation}
	\CI{ }\left(p,q\right) = -\min_{0\leq s \leq 1} \log \displaystyle\sum_{i=1}^n  p_i^{1-s}q_i^s.
\end{equation}
It is useful to define yet another Chernoff distance on quantum states, for an even more restricted class of measurements than LOCC. Let $(M, \1 - M)$ be the optimal single-copy LOCC POVM. $\CI{SC}(\rho_1,\rho_2;p)$ is the classical Chernoff distance between the probability distributions on the outcome of this measurement when it is performed on $\rho_1$ or $\rho_2$. (Outside the bipartite setting this notion was considered before by Kargin~\cite{kargin}.) If we write
\begin{equation}
	p_{j1} = \tr\left( M \rho_j \right),\  
	p_{j2} = \tr\bigl( (\1 - M) \rho_j \bigr),
\end{equation}
we can summarize the relationships between Chernoff distances we have defined as follows:

\begin{equation}
	- \min_{0\leq s \leq 1} \log \sum_{i=1}^{2} p_{1i}^{1-s} p_{2i}^{s} = \CI{SC} \leq \CI{LOCC} \leq \CI{SEP} \leq \CI{PPT} \leq \CI{ALL} = - \min_{0\leq s \leq 1} \log \tr \rho_1^{1-s} \rho_2^{s}
\end{equation}


Before proceeding with our main new results, we would like to make some general remarks about these quantities and describe some of the existing knowledge about them.
One striking difference between global and local state discrimination can be seen in the effect of adding an ancilla. In the global case, this has no effect on our ability to distinguish between states, asymptotically or otherwise. That is, for any state $\tau$
\begin{equation}
	\Perr{ALL}\left(\rho_1,\rho_2;p\right) = \Perr{ALL}\left(\rho_1\ox\tau,\rho_2\ox\tau;p\right),\ 
    \CI{ALL}\left(\rho_1,\rho_2;p\right) = \CI{ALL}\left(\rho_1\ox\tau,\rho_2\ox\tau;p\right).
\end{equation}
This is hardly surprising when one considers that the addition of any ancilla state is subsumed by the POVM formalism in the global case.
In cases where our ability to distinguish between two states (of a $d \times d$ system, let's say) is worsened by restriction to LOCC, then we will indeed be helped by the provision of a $d \times d$ maximally entangled ancilla: by using it to teleport Alice's half to Bob (say), we have restored the ability to make global measurements and will be able to decrease the error probability accordingly.
It is not always the case that the restriction to LOCC will impair our performance however. It was shown by Walgate \emph{et al.} \cite{walgate} (and generalized to non-orthogonal states by Virmani \emph{et al.} \cite{virmani}) that LOCC can do just as well in distinguishing between two pure states as a global measurement can.
\begin{equation}
	\Perr{ALL}\left(\ketbra{\psi}{\psi},\ketbra{\phi}{\phi};p\right) = \Perr{LOCC}\left(\ketbra{\psi}{\psi},\ketbra{\phi}{\phi};p\right).
\end{equation}
Naturally, the corresponding Chernoff distances are also equal when both states are pure.
Recently, Nathanson~\cite{nathanson-unpub} has generalized this to
the case of discriminating a mixed state from a pure state. He finds that
under certain conditions on the fidelity of the states and the Schmidt coefficients
of the pure state, $\CI{LOCC}(\rho_1,\rho_2) = \CI{ALL}(\rho_1,\rho_2)$, even
though the single-copy error probabilities may differ.


From our perspective, it is more interesting to look at pairs of states where the LOCC constraint reduces our ability to distinguish them. In this paper we discuss an example of such a case. Let $\s_d$ and $\a_d$ denote the completely symmetric and completely anti-symmetric Werner states in $d \times d$ dimensions, respectively (when $d$ is a power of two, these are the states used by DiVincenzo \emph{et al.}~\cite{data-hiding} for ``data hiding''; see also \cite{Werner}). In this paper we calculate the Chernoff distance between these states, $\CI{LOCC}\left(\s_d,\a_d\right)$, and to do so, we actually give an expression for $\Perr{LOCC}\left(\s_d^{\ox n},\a_d^{\ox n};p\right)$.

The rest of this paper is organized as follows: In the next section we present an LOCC protocol which puts an upper bound on $\Perr{LOCC}\left(\s_d^{\ox n},\a_d^{\ox n};p\right)$. In section~\ref{sec:ppt}, we formulate the minimization of the error which can by achieved by PPT operations as a linear program, and by solving the dual program show that the LOCC upper bound is also a lower bound on $\Perr{PPT}\left(\s_d^{\ox n},\a_d^{\ox n};p\right)$ and hence on $\Perr{LOCC}\left(\s_d^{\ox n},\a_d^{\ox n};p\right)$, thus proving the optimality of our LOCC protocol, and allowing us to calculate the Chernoff distance. In section~\ref{sec:biasbound}, we prove a lower bound on $\bias{SEP}\left(\rho_1,\rho_2;p\right)$ in terms of $\bias{ALL}\left(\rho_1,\rho_2;p\right)$, after which we conclude.

To describe asymptotic behaviours we will use `Big-O' notation (including $\Theta, \Omega$ and $\sim$). If $X$ is an operator on a bipartite Hilbert space $\mathcal{H}_A \ox \mathcal{H}_B$, we use $X\pt$ to denote its partial transpose, which is defined (for some orthonormal product basis $\{\ket{i}_A \ox \ket{j}_B\}$) by

\begin{equation}
	\ket{i}_A \ox \ket{j}_B \bra{k}_A \ox \bra{l}_B \pt = \ket{i}_A \ox \ket{l}_B \bra{k}_A \ox \bra{j}_B .
\end{equation}


\section{LOCC Discrimination Protocol}
\label{sec:locc}

\begin{proposition}
\label{prop:locc}
 There is an LOCC protocol (requiring only one-way communication) which demonstrates that $\Perr{LOCC}\left(\s_d^{\ox n},\a_d^{\ox n};p\right) \leq \min\left(p\left(\frac{d-1}{d+1}\right)^n, 1-p\right)$.
\end{proposition}
\begin{proof}
Alice and Bob take each copy in turn and measure in the computational basis. They share their results. If they recorded different results for every copy then they guess that they have the anti-symmetric state. Otherwise, they have obtained the same result for at least one state and they know with certainty that they share the symmetric state.

For a single copy, the POVM implemented by this measurement is
\begin{equation}
\Bigg\{G_{d} = \sum_{i\neq j}^{d-1}\ketbra{ij}{ij}, \openone - G_{d} = \sum_{i=0}^{d-1}\ketbra{ii}{ii} \Bigg\}.
\end{equation}

Because the states to be distinguished are both $U\otimes U$-invariant,
it is convenient to apply the twirl operation to the two operators in the POVM and it also emphasizes the symmetry of the states that are to be distinguished. After doing so we have
the following single-copy POVM of equal performance:
\begin{equation}
\Big\{M_{d} = \frac{d-1}{d+1}\Pi_s + \Pi_a, \openone -M_{d} = \frac{2}{d+1} \Pi_s \Big\},
\end{equation}
where $\Pi_s$ and $\Pi_a$ are the projections  onto the symmetric and anti-symmetric subspaces, respectively.
The POVM element $M_{d}$ corresponds to Alice and Bob having different measurement outcomes on a single copy. For $n$ copies the POVM is
\begin{equation}
\{M_{d}^{\ox n}, \openone - M_{d}^{\ox n}\},
\end{equation}
since $M_{d}^{\ox n}$ corresponds to Alice and Bob getting different outcomes for every copy they measure.
Let $A_{k}$ denote the sum of all elements of $\{\Pi_s, \Pi_a\}^{\ox n}$ which have $k$ copies of $\Pi_a$. Expanding in terms of the $n+1$ orthogonal projection operators $\{A_0,\ldots,A_n\}$, we find that
\begin{equation}\label{povmExpansion}
M_{d}^{\ox n}=\sum_{k=0}^n \left(\frac{d-1}{d+1}\right)^{n-k} A_{k}.
\end{equation}

\begin{equation}
\label{errorProb1}
  \Perr{ } = p \tr \left(M_d^{\ox n} \s_d^{\ox n}\right) + (1-p) \tr \Bigl(\left(\openone - M_{d}^{\ox n}\right) \a_d^{\ox n}\Bigr),
\end{equation}
where the first term is the probability that Alice and Bob have the symmetric state and mistake it for the anti-symmetric state and the second term is the probability that they share the anti-symmetric and mistake it for the symmetric state.

Substituting \eref{povmExpansion} into \eref{errorProb1} and using the fact that $\s_d^{\ox n} \propto A_0$ and $\a_d^{\ox n} \propto A_n$, we obtain
\begin{equation}\label{errorProbMain}
P_{\rm err} = p \tr\left(\left(\frac{d-1}{d+1}\right)^n A_0 \s_d^{\ox n}\right) + (1-p) \tr \left(\left(1-A_n\right) \a_d^{\ox n}\right) = p \left(\frac{d-1}{d+1}\right)^n .
\end{equation}
If $P_{\rm err} > 1 - p$ then we will do better to simply guess that we have the symmetric state all the time. Adding this proviso to our strategy, we obtain the desired result.
\end{proof}

\begin{remark}
We note that the second term in the expression for the error probability is zero, meaning that all the error is due to the case where the symmetric state is mistaken for the anti-symmetric state. This is just what we would expect given that our protocol reports that we have a symmetric state only when it is \emph{certain} that we have one.
\end{remark}

We shall now show that \eref{errorProbMain} is the optimum error probability that can be achieved using LOCC by showing that it is the best that can be achieved even if we use the larger class of measurements that can be implemented using PPT preserving operations.


\section{Optimal PPT preserving POVM}
\label{sec:ppt}
We shall first formulate the minimisation of the error probability over PPT preserving POVMs~\cite{rains} as linear programming problem (see \cite{LP}, for instance) by taking advantage of the symmetries of the states we wish to distinguish. We will then show that there is a solution to the dual linear program which lower bounds the error probability to exactly that achieved by the LOCC procedure given above.

The states $\a_d^{\ox n}$ and $\s_d^{\ox n}$ are invariant under permutations of the copies and under biunitary transformations of the individual copies. We can assume therefore that our two POVM elements have the same symmetries (this is a trick that was used before in~\cite{E_re} to solve a relative entropy minimisation problem).
The most general operator with these symmetries is a linear combination of the $n+1$ operators $A_k$ which we defined above, so we write our POVM as:

\begin{equation}
\left\{ \sum_{k=0}^n x_k A_k, \sum_{k=0}^n (1 - x_k) A_k \right\}.
\end{equation}
The probability of error is given by
\begin{equation}\label{errorPPT}
P_{\rm err} = p \tr \left(\sum_{k=0}^n x_k A_k \s_d^{\ox n}\right) + (1-p) \tr \left(\sum_{k=0}^n (1 - x_k) A_k \a_d^{\ox n}\right) = (1-p) + p \left(x_0-\frac{1-p}{p}x_n\right) .
\end{equation}
The constraints
\begin{eqnarray}
x_k \geq 0 & \textrm{ for } k = 0,\ldots, n, \\
x_k \leq 1 & \textrm{ for } k = 0,\ldots, n
\end{eqnarray}
are necessary and sufficient to ensure that the two operators do in fact comprise a POVM.

The partial transpose of the flip operator $F$ is equal to $d\Phi_d$, where
$\Phi_d = \frac{1}{d}\sum_{i,j=0}^{d-1}\ketbra{ii}{jj}$ is the maximally entangled state.
Since $\Pi_s = (\openone+F)/2$ and $\Pi_a = (\openone-F)/2$, we have
\begin{eqnarray}
\Pi_s\pt = \frac{1}{2}\left(\openone + d\Phi_d\right)
         = \frac{1}{2}\bigl(\left(\openone-\Phi_d\right) + \left(1+d\right)\Phi_d\bigr),\\
\Pi_a\pt = \frac{1}{2}\left(\openone - d\Phi_d\right)
         = \frac{1}{2}\bigl(\left(\openone-\Phi_d\right) + \left(1-d\right)\Phi_d\bigr),
\end{eqnarray}
so the operators $A_k\pt$ can be written as linear combinations of operators from the set of $2^n$ orthogonal operators $\left\{(\openone-\Phi_d), \Phi_d\right\}^{\ox n}$.

Let $S_k^n$ denote the subset of strings in $\left\{0,1\right\}^N$ which have exactly $k$ ones. Then,
\begin{equation}
\begin{aligned}
A_k\pt &= 2^{-n}\sum_{v \in S_k^n}\OX_{i=1}^n \bigl(\left(\openone-\Phi_d\right)+\left(1+\left(-1\right)^{v_i}d\right)\Phi_d\bigr) 
\\
&= 2^{-n} \sum_{l=0}^n\sum_{0\leq j \leq l,k} \binom{n-l}{k-j} \binom{l}{j} (1+d)^j(1-d)^{l-j} T_l,
\end{aligned}
\end{equation}
where $T_l$ is the sum over all elements of $\left\{\left(\openone-\Phi_d\right), \Phi_d\right\}^{\ox n}$ which have $l$ copies of $\Phi_d$.

A POVM is PPT preserving if and only if all of the operators that comprise it have positive partial transpose~\cite{rains}. A necessary and sufficient condition for the POVM to be PPT preserving is therefore given by the following inequalities
\begin{eqnarray}
\sum_{k=0}^{n}x_k\sum_{0 \leq j \leq l,k}\binom{n-l}{k-j}\binom{l}{j}\left(1-d\right)^j\left(1+d\right)^{l-j} \geq 0 & \textrm{ for } l = 0,\ldots, n, \\
\sum_{k=0}^{n}(1-x_k)\sum_{0 \leq j \leq l,k}\binom{n-l}{k-j}\binom{l}{j}\left(1-d\right)^j\left(1+d\right)^{l-j} \geq 0 & \textrm{ for } l = 0,\ldots, n.
\end{eqnarray}

Let $Q$ be an $(n+1)\times (n+1)$ matrix with elements
\begin{equation}
Q_{lk} = \sum_{0 \leq j \leq l,k}\binom{n-l}{k-j}\binom{l}{j}\left(1-d\right)^j\left(1+d\right)^{l-j}.
\end{equation}

We note that
\begin{equation}
\begin{aligned}
\sum_{k=0}^{n}Q_{lk} &= \left(1+d\right)^l \sum_{m=0}^{n-l} \binom{n-l}{m} \sum_{j=0}^{l}\binom{l}{j}\left(\frac{1-d}{1+d}\right)^j\\
&= \left(1+d\right)^l \left(1+\frac{1-d}{1+d}\right)^l \sum_{m=0}^{n-l}\binom{n-l}{m}\\
&= \left(1+d\right)^l \left(\frac{2}{1+d}\right)^l 2^{n-l} = 2^n.
\end{aligned}
\end{equation}
Defining the vectors $c$ and $b$ as follows
\begin{eqnarray}
c_i &=& \delta_{0i} - \frac{1-p}{p} \delta_{ni}, \\
b_i &=& \left\{ \begin{array}{lll}
0 &\textrm{ for } i = 0,\ldots, n, \\
-2^n &\textrm{ for } i = n+1,\ldots ,2n+1,\\
-1 &\textrm{ for } i = 2n+2,\ldots ,3n+2,
\end{array} \right.
\end{eqnarray} 
we can write the optimisation in standard linear programming form
\begin{eqnarray}\label{primalLP}
\min_x \{ c^T\cdot x | P\cdot x \geq b , x \geq 0 \} \textrm{ where } P = \left( \begin{array}{c}
Q \\
-Q \\
-\openone
\end{array} \right).
\end{eqnarray}

Writing \eref{errorPPT} in terms of the objective function $c^T\cdot x$, we see that the POVM corresponding the vector $x$ has error probability
\begin{equation}
P_{\rm err}(x) = (1-p) + p c^T \cdot x.
\end{equation}

\begin{proposition}
\label{prop:ppt}
The probability of error for a PPT preserving POVM to distinguish $\s_d^{\ox n}$ and $\a_d^{\ox n}$ with prior probabilities $p$ and $1-p$, $\Perr{PPT}\left(\s_d^{\ox n},\a_d^{\ox n}; p\right)$, is bounded below by $\min\left(p \left(\frac{d-1}{d+1}\right)^n, 1-p \right)$.
\end{proposition}
\begin{proof}
The linear program dual to \eref{primalLP} is just
\begin{equation}
\max_y \{ b^T\cdot y | P^T\cdot y \leq c , y \geq 0 \}.
\end{equation}

Indeed, the duality of linear programs tells that for any primal feasible point $x$ and any dual feasible point $y$
\begin{equation}
c^T\cdot x \geq b^T\cdot y,
\end{equation}
so any dual feasible point $y$ gives us a lower bound on the error probability:
\begin{equation} \label{errorBound}
\Perr{PPT}\left(\s_d^{\ox n},\a_d^{\ox n};p\right) \geq (1-p) + p b^T \cdot y.
\end{equation}

It is convenient to write $y$ as the direct sum of three $(n+1)$-dimensional vectors $y = u \oplus v \oplus w$ so that we can rewrite the dual program as
\begin{eqnarray}\label{dualLP}
\max_y \left\{ - 2^n \sum_{i=0}^n v_i - \sum_{i=0}^n w_i \Big| u\geq 0, v\geq 0, w\geq 0, Q^T\cdot u - Q^T\cdot v - w \leq c \right\}.
\end{eqnarray}

Consider the point $y^{\ast}$ = $u^{\ast}\oplus v^{\ast}\oplus w^{\ast}$ defined by
\begin{eqnarray}
u^{\ast}_i &=& \binom{n}{i}\frac{(d-1)^{n-i}\bigl((d+1)^i - (1-d)^i\bigr)}{(2d)^n (d+1)^i}, \\
v^{\ast}_i &=& 0, \\
w^{\ast}_i &=& \left\{ \begin{array}{ll}
0 &\textrm{ for } i = 0,\ldots ,n-1, \\
\max \left(\frac{1-p}{p}-(\frac{d-1}{d+1})^n,0 \right) &\textrm{ for } i = n.
\end{array} \right.
\end{eqnarray}

We show that the point $y^{\ast}$ is dual feasible in Appendix A. The dual objective function at this point is
\begin{equation}
-2^n \sum_{i=0}^n v^{\ast}_i - \sum_{i=0}^n w^{\ast}_i = - w^{\ast}_n = \min\left(\left(\frac{d-1}{d+1}\right)^n-\frac{1-p}{p},0\right),
\end{equation}
so, substituting $y^{\ast}$ into \eref{errorBound}, we obtain the bound:
\begin{equation}
\Perr{PPT}\left(\s_d^{\ox n},\a_d^{\ox n};p\right) \geq \min\left(p\left(\frac{d-1}{d+1}\right)^n,1-p\right).
\end{equation}
\end{proof}

\begin{corollary}
	Substituting the results of Proposition \ref{prop:locc} and Proposition \ref{prop:ppt} into \eref{errorinequalities}, we have shown that
\begin{equation}
	\Perr{PPT}\left(\s_d^{\ox n},\a_d^{\ox n};p\right) = \Perr{SEP}\left(\s_d^{\ox n},\a_d^{\ox n};p\right) = \Perr{LOCC}\left(\s_d^{\ox n},\a_d^{\ox n};p\right) = \min\left(p\left(\frac{d-1}{d+1}\right)^n,1-p\right).
\end{equation}
\end{corollary}
Substituting into the definition of the Chernoff information for each class of operations and noting that each copy is measured separately in the optimal strategy, we obtain our main result:
\begin{theorem}
	Whenever $0<p<1$, we have
\begin{equation}
	\CI{PPT}\left(\s_d,\a_d\right) = \CI{SEP}\left(\s_d,\a_d\right) = \CI{LOCC}\left(\s_d,\a_d\right) = \CI{SC}\left(\s_d,\a_d\right) = \log \frac{d+1}{d-1} \sim \frac{2\log e}{d-1}.
\end{equation}
\end{theorem}


\section{A Lower Bound on Bias for\protect\\ Single-Copy Separable Measurements}
\label{sec:biasbound}
The fact that $\CI{LOCC}\left(\s_d,\a_d\right) = \CI{SC}\left(\s_d,\a_d\right)$ shows that our ability to distinguish the extremal Werner states cannot be improved by measurements which are entangled across copies. This is the least favorable many-copy behaviour possible. It would be interesting to know if the single-copy error probability for these states also has the worst kind of scaling with dimension. In terms of bias, we have shown that
\begin{equation}
	\frac{\bias{LOCC}\left(\s_d,\a_d;p\right)}{\bias{ALL}\left(\s_d,\a_d;p\right)}
            	    = \Theta\left(\frac{1}{d}\right).
\end{equation}
Is $1/d$ an asymptotic lower bound whatever states we choose? If we relax the LOCC constraint and allow separable operations then we can show that it is.
\begin{proposition}
	If $\rho_1$ and $\rho_2$ are bipartite states on a system of overall dimension $D$, then 
\begin{equation}
	\bias{SEP}\left(\rho_1,\rho_2;p\right) \geq \frac{1}{2\sqrt{D}}\bias{ALL}\left(\rho_1,\rho_2;p\right).
\end{equation}
\end{proposition}
\begin{proof}
We know that the optimal error probability for global measurements is given by the Holevo-Helstrom POVM, the elements of which are generally not even PPT. It was shown by Barnum and Gurvits \cite{sepball} that every Hermitian operator in the ball centred on the identity, with radius one in the Hilbert-Schmidt norm is separable. If we add to each element of the Holevo-Helstrom POVM the minimum amount of the identity operator necessary to put the resulting operator inside this ball, and normalize the POVM, we obtain the \emph{separable} POVM

\begin{equation}
	\left\{ \frac{1}{2}\left( \1 + \frac{M}{\|M\|_2} \right), \frac{1}{2}\left( \1 - \frac{M}{\|M\|_2} \right) \right\}
\end{equation}
where $M$ is the projector onto the support of the positive part of $(1-p)\rho_2 - p\rho_1$ if $p \leq 1/2$ (and minus one times the projector onto the support of the negative part otherwise). This POVM yields the error probability
\begin{equation}
	\Perr{ } = \frac{1}{2}\left(1 - \frac{1}{2\|M\|_2}\bigl(|1-2p| + \|(1-p)\rho_2 - p\rho_1\|_1\bigr)\right).
\end{equation}
Using the fact that $\|M\|_2 \leq \sqrt{D}$, we get the bound

\begin{equation}
    \label{eq:bias-chain}
	\|(1-p)\rho_2 - p\rho_1\|_1 =    \bias{ALL}
	                            \geq \bias{PPT} 
	                            \geq \bias{SEP} 
	                            \geq \frac{1}{2\sqrt{D}}\|(1-p)\rho_2 - p\rho_1\|_1
	                            =     \frac{1}{2\sqrt{D}}\bias{ALL}.
\end{equation}
\end{proof}

So, for states of a $d \times d$ system: $\bias{SEP}/\bias{ALL} \in \Omega(1/d)$. This result, combined with our result for the the data hiding states, leads us to conjecture that
\begin{conjecture}
	\label{conj:locc-bias}
	For states on a $d \times d$ system,
    \begin{equation}
	  \frac{\bias{LOCC}}{\bias{ALL}} \geq \Omega\left(\frac{1}{d}\right).
    \end{equation}
\end{conjecture}

\medskip
To put the insights and conjecture above into a different and wider perspective,
let us look at the biases $\bias{X}$ for the particular value $p=\frac{1}{2}$:
\begin{equation}
  \bias{X}(\rho_1,\rho_2) := \bias{X}\left(\rho_1,\rho_2;\frac{1}{2}\right),
\end{equation}
for which, by definition, it is clear that it is symmetric:
$\bias{X}(\rho_1,\rho_2) = \bias{X}(\rho_2,\rho_1)$. Furthermore, for
all the classes $X$ considered in the introduction,
$\bias{X}(\rho_1,\rho_2) = 0$ if and only if $\rho_1 = \rho_2$.
Indeed, the $\bias{X}$ are all metrics, as they obey the triangle inequality:
$\bias{X}(\rho_1,\rho_3) \leq \bias{X}(\rho_1,\rho_2) + \bias{X}(\rho_2,\rho_3)$
for any states $\rho_1$, $\rho_2$ and $\rho_3$.
To be more precise, they derive from operator norms $\|\cdot\|_X$, defined on trace-free hermitian operators:
\begin{equation}\label{opnormdfn}
  \bias{X}(\rho_1,\rho_2) = \left\| \frac{1}{2}(\rho_1-\rho_2) \right\|_X,
  \text{ with }
  \| M \|_X = \sup_{\text{POVM }(M_i)_i \in \text{X}} \ \sum_i |\tr M M_i|,
\end{equation}
We note that the supremum in \eref{opnormdfn} is always attained by a POVM with two elements (one with $\tr (M M_1) \geq 0$ and the other with $\tr (M M_2) = -\tr (M M_1) \leq 0$).

For example by Helstrom's theorem~\cite{helstrom},
$\bias{X}(\rho_1,\rho_2) = \left\| \frac{1}{2}(\rho_1-\rho_2) \right\|_1$,
so $\|\cdot\|_{\text{ALL}} = \|\cdot\|_1$.

Of course, all norms on finite-dimensional spaces are equivalent up
to constant factors. Eq.~(\ref{eq:bias-chain}) translates into the ordering of norms
\begin{equation}
  \label{eq:norm-chain}
  \| M \|_1 =    \| M \|_{\text{ALL}}
                 \geq \| M \|_{\text{PPT}}
                 \geq \| M \|_{\text{SEP}}
                 \geq \sqrt{\frac{1}{D}} \| M \|_{\text{ALL}},
\end{equation}
and Conjecture~\ref{conj:locc-bias} can be expressed as
$\| M \|_{\text{LOCC}} \geq \Omega\left(\frac{1}{d}\right) \| M \|_{\text{ALL}}$
for $d \times d$ systems. Note that the existence of data hiding states
implies that this would be essentially best possible, as for
$M = \frac{1}{2}(\alpha_d-\sigma_d)$,
\begin{equation}
  \| M \|_{\text{LOCC}} \leq \| M \|_{\text{SEP}}
                             \leq \| M \|_{\text{PPT}}
                             =    \frac{2}{d+1} \| M \|_{\text{ALL}}.
\end{equation}

\section{Discussion}
We have calculated the Chernoff distance between the extremal $d \times d$ Werner states, under the constraint of LOCC operations, for all values of $d$. This is the first time the LOCC Chernoff distance has been calculated for states where it differs from the unconstrained Chernoff distance. In this case, we have also been able to calculate the smallest error probability that can be achieved by LOCC for any finite number of copies. The solution has at least two remarkable features: First, the error probability is -- up to constant factors -- equal to the $n$-th power of the single-copy error probability, showing that in a sense $n$ copies don't give disproportionate advantage over one copy, in this case. Secondly, even the optimal $n$-copy measurement reflects this structurally; namely, it can be implemented by measuring the single-copy optimal POVM $n$ times, followed by a trivial classical post-processing. As discussed in the introduction, this is a ``worst-case'' strategy for many copies. Both of these properties distinguish the solution from what is to be expected in the quantum Chernoff problem: e.g., discriminating two (non-orthogonal) pure states has a very simple optimal strategy, but for $n$ copies (which is also a problem of discriminating two pure states) this strategy is highly collective over the $n$ systems. Also, in general, even classically, the error probability shows only an asymptotically exponential decay, but here it is exactly exponential.
\par
Our result also leads to a number of further questions. An extension of the work which we are currently considering is to see if we can find Chernoff bounds for the discrimination of pairs of general Werner states. Preliminary and ongoing investigations suggest that some interesting effects occur when at least one state is non-extremal. Also, as discussed above, it would be interesting to know how close to ``worst possible'' is our example in terms of comparing LOCC to unrestricted measurements? That is, we would like to resolve our Conjecture~\ref{conj:locc-bias} on the single-copy LOCC bias.

\begin{acknowledgments}
WM acknowledges support from the U.K. EPSRC; AW was supported through an Advanced Research Fellowship of the U.K. EPSRC, the EPSRC's ``QIP IRC'', and the European Commission IP ``QAP''. The Centre for Quantum Technologies is funded by the Singapore Ministry of Education
and the National Research Foundation as part of the
Research Centres of Excellence programme.

The authors would like to acknowledge useful discussions with Keiji Matsumoto, Chris King and Michael Nathanson and to thank Aram Harrow for a stimulating conversation on the design of the optimally discriminating POVM.
\end{acknowledgments}

\appendix
\section{Proof of dual feasibility}
\label{sec:dualproof}
We note that $u^{\ast}_i \geq 0$ for $i = 0,\ldots ,n$:
\begin{equation}
\begin{aligned}
u^{\ast}_k &= (2d)^{-n}\binom{n}{k}\left(\left(d-1\right)^{n-k}-(-1)^k \frac{\left(d-1\right)^n}{\left(d+1\right)^k}\right) \\
&\geq (2d)^{-n}\binom{n}{k}\left(\left(d-1\right)^{n-k}- \frac{\left(d-1\right)^n}{\left(d+1\right)^k}\right)\\
&= \binom{n}{k}\left(\frac{d-1}{2d}\right)^{n}\left(\frac{1}{\left(d-1\right)^{k}}-\frac{1}{\left(d+1\right)^{k}}\right) \geq 0.
\end{aligned}
\end{equation}
It is obvious that $v^{\ast} \geq 0$ and $w^{\ast} \geq 0$, so the first three inequalities of \eref{dualLP} are satisfied.

We now show that the remaining inequality,
\begin{equation}\label{Qconstraint}
Q^T\cdot u - Q^T\cdot v - w \leq c,
\end{equation}
 is also satisfied:
\begin{equation}\label{Qu}
\begin{aligned}
( Q^T\cdot u^{\ast} )_k &=
\frac{(d-1)^n}{(2d)^n}\sum_{l=0}^{n} \sum_{0 \leq j \leq l,k}\binom{n-l}{k-j}\binom{l}{j}\binom{n}{l}\left(1-d\right)^j\left(1+d\right)^{l-j} \frac{(d+1)^l - (1-d)^l}{(d-1)^l (d+1)^l}\\
&= s_1(d,n;l)-s_2(d,n;l),
\end{aligned}
\end{equation}
where
\begin{equation}\label{s1}
\begin{aligned}
s_1(d,n;k) &= \frac{(d-1)^n}{(2d)^n}\sum_{l=0}^{n} \sum_{0 \leq j \leq l,k}\binom{n-l}{k-j}\binom{l}{j}\binom{n}{l}\left(1-d\right)^j\left(1+d\right)^{l-j} \frac{(d+1)^l}{(d-1)^l (d+1)^l}\\ 
&= \frac{(d-1)^n}{(2d)^n}\sum_{l=0}^{n} \sum_{0 \leq j \leq l,k} \binom{n-l}{k-j}\binom{l}{j}\binom{n}{l} (-1)^j\left(\frac{d+1}{d-1}\right)^{l-j},
\end{aligned}
\end{equation}
\begin{equation}\label{s2}
\begin{aligned}
s_2(d,n;k) &= \frac{(d-1)^n}{(2d)^n}\sum_{l=0}^{n} \sum_{0 \leq j \leq l,k}\binom{n-l}{k-j}\binom{l}{j}\binom{n}{l}\left(1-d\right)^j\left(1+d\right)^{l-j} \frac{(1-d)^l}{(d-1)^l (d+1)^l}\\
&= \frac{(d-1)^n}{(2d)^n}\sum_{l=0}^{n} \sum_{0 \leq j \leq l,k}\binom{n-l}{k-j}\binom{l}{j}\binom{n}{l}(-1)^{j+l}\left(\frac{d-1}{d+1}\right)^j.
\end{aligned}
\end{equation}
Defining $m = l - j$ we can rewrite the first sum \eref{s1} as
\begin{equation}
\begin{aligned}
s_1(d,n;k) &= \frac{(d-1)^n}{(2d)^n}\sum_{m=0}^{n-k} \sum_{j =0}^{k}\binom{n-(m+j)}{k-j}\binom{m+j}{j}\binom{n}{m+j} (-1)^j\left(\frac{d+1}{d-1}\right)^m\\
&=  \frac{(d-1)^n}{(2d)^n}\sum_{m=0}^{n-k} \sum_{j =0}^{k}\frac{n!}{(k-j)!(n-(m+k))!m!j!} (-1)^j\left(\frac{d+1}{d-1}\right)^m\\
&= \frac{(d-1)^n}{(2d)^n}\sum_{m=0}^{n-k} \sum_{j =0}^{k}\frac{n!}{(n-m)!m!}\frac{(n-m)!}{((n-m)-k)!k!}\frac{k!}{(k-j)!j!} (-1)^j\left(\frac{d+1}{d-1}\right)^m\\
&= \frac{(d-1)^n}{(2d)^n}\sum_{m=0}^{n-k} \binom{n}{m}\binom{n-m}{k}\left(\frac{d+1}{d-1}\right)^m \sum_{j =0}^{k} \binom{k}{j} (-1)^j.
\end{aligned}
\end{equation}
The sum over $j$ is $0$ except when $k=0$, so
\begin{equation}\label{s1simp}
s_1(d,n;k) = \delta_{0k}\frac{(d-1)^n}{(2d)^n}\sum_{m=0}^{n}\binom{n}{m}\left(\frac{d+1}{d-1}\right)^m = \delta_{0k}\frac{(d-1)^n}{(2d)^n}\left(1+\frac{d+1}{d-1}\right)^n = \delta_{0k} .
\end{equation}
Making the same change of variables ($m = l - j$) in \eref{s2}, we obtain
\begin{equation}
\begin{aligned}\label{s2simp}
s_2(d,n;k) &= \frac{(d-1)^n}{(2d)^n} \sum_{j=0}^{k} \sum_{l=j}^{n+j-k}\binom{n}{l}\binom{n-l}{k-j}\binom{l}{j}(-1)^{j+l}\left(\frac{d-1}{d+1}\right)^j\\
 &= \frac{(d-1)^n}{(2d)^n} \sum_{j=0}^{k} \sum_{m=0}^{n-k}\binom{n}{m+j}\binom{n-(m+j)}{k-j}\binom{m+j}{j}(-1)^{j}(-1)^{m+j}\left(\frac{d-1}{d+1}\right)^j\\
&= \frac{(d-1)^n}{(2d)^n} \sum_{j=0}^{k} \sum_{m=0}^{n-k}\frac{n!}{(k-j)!(n-(m+k))!m!j!}(-1)^{2j}(-1)^{m}\left(\frac{d-1}{d+1}\right)^j\\
&= \frac{(d-1)^n}{(2d)^n} \sum_{j=0}^{k} \sum_{m=0}^{n-k}\frac{n!}{(n-k)!k!}\frac{(n-k)!}{((n-k)-m)!m!}\frac{k!}{(k-j)!j!}(-1)^{m}\left(\frac{d-1}{d+1}\right)^j\\
&= \frac{(d-1)^n}{(2d)^n}  \binom{n}{k}\sum_{m=0}^{n-k} \binom{n-k}{m}(-1)^{m}\sum_{j=0}^{k}\binom{k}{j}\left(\frac{d-1}{d+1}\right)^j\\
&= \delta_{nk}\frac{(d-1)^n}{(2d)^n}  \binom{n}{k} \left(1+\frac{d-1}{d+1}\right)^n = \delta_{nk}\left(\frac{d-1}{d+1}\right)^n .
\end{aligned}
\end{equation}
Substituting \eref{s1simp} and \eref{s2simp} into \eref{Qu} we find that $(Q^T\cdot u^{\ast})_k = \delta_{0k} - \delta_{nk}\left(\frac{d-1}{d+1}\right)^n$, so the constraint \eref{Qconstraint} is satisfied:
\begin{equation}
(Q^T\cdot u^{\ast} - Q^T\cdot v^{\ast} - w^{\ast})_k = \delta_{0k} - \delta_{nk}\left(\frac{d-1}{d+1}\right)^n - \max\left(r- \left(\frac{d-1}{d+1}\right)^n, 0\right)\delta_{nk} \leq c_k .
\end{equation}

\end{document}